# Forming the Moon from terrestrial silicate-rich material

R.J. de Meijer[1,2], V.F. Anisichkin[3], W. van Westrenen[4,*]

[1]Stichting EARTH, de Weehorst, 9321 XS 2, Peize, the Netherlands

[2]Department of Physics, University of the Western Cape, Private Bag X17, Bellville 7535, South Africa

[3]Lavrentyev Institute of Hydrodynamics, Siberian Branch of Russian Academy of Sciences, Novosibirsk, Russia

[4]Faculty of Earth and Life Sciences, VU University Amsterdam, De Boelelaan 1085, 1081 HV Amsterdam, the Netherlands

[*]Corresponding author. E-mail: w.van.westrenen@vu.nl

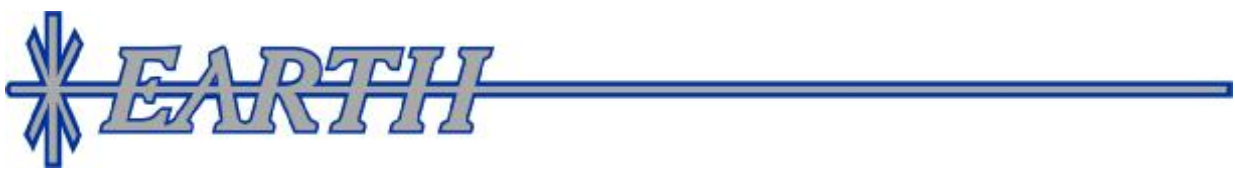

"The history of science has included many premature rejections of ideas which were later adopted. As new data become available it is well to reconsider the grounds on which former hypotheses were rejected"
– Donald U. Wise (1963)

**Abstract**

Recent high-precision measurements of the isotopic composition of lunar rocks demonstrate that the bulk silicate Earth and the Moon show an unexpectedly high degree of similarity. This is inconsistent with one of the primary results of classic dynamical simulations of the widely accepted giant impact model for the formation of the Moon, namely that most of the mass of the Moon originates from the impactor, not Earth.

Resolution of this discrepancy without changing the main premises of the giant impact model requires total isotopic homogenisation of Earth and impactor material after the impact for a wide range of elements including oxygen, silicon, potassium, titanium, neodymium, and tungsten. Isotopic exchange between partially molten and vaporised Earth and Moon shortly after the impact has been invoked to explain the identical oxygen isotopic composition of Moon and Earth but the effectiveness and dynamics of this process are contested. Even if this process could explain the O isotope similarity, it is unlikely to work for the much heavier, refractory elements. Given the increasing uncertainty surrounding the giant impact model in light of these geochemical data, alternative hypotheses for lunar formation should be explored.

In this paper, we revisit the hypothesis that the Moon was formed directly from terrestrial mantle material, as first proposed in the 'fission' hypothesis (Darwin, G.H., 1879. On the bodily tides of viscous and semi-elastic spheroids, and on the ocean tides upon a yielding nucleus. Phil. Trans. Roy. Soc. (London) 170, 1-35). We show that the dynamics of this scenario requires on the order of $10^{29}$-$10^{30}$ J almost instantaneously generated additional energy if the angular momentum of the proto-Earth was similar to that of the Earth-Moon system today. The only known source for this additional energy is nuclear fission. We show that it is feasible to form the Moon through the ejection of terrestrial silicate material triggered by a nuclear explosion at Earth's core-mantle boundary (CMB), causing a shock wave propagating through the Earth. Hydrodynamic modelling of this scenario shows that a shock wave created by rapidly expanding plasma resulting from the explosion disrupts and expels overlying mantle and crust material. Our hypothesis straightforwardly explains the identical isotopic composition of Earth and Moon for both lighter (oxygen, silicon, potassium) and heavier (chromium, titanium, neodymium and tungsten) elements. It is also consistent with proposed Earth-like water abundances in the early Moon, with the angular momentum and energy of the present-day Earth-Moon system, and with the early formation of a 'hidden reservoir' at Earth's CMB that is not present in the Moon.

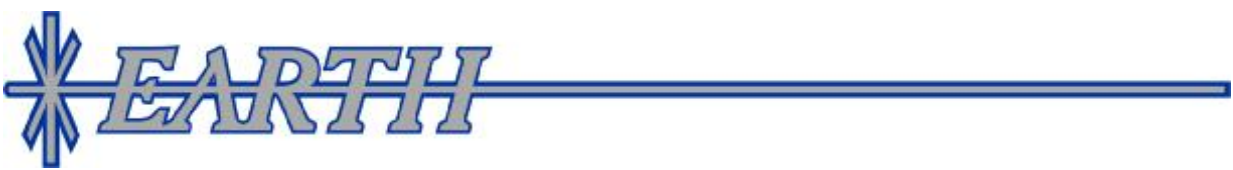

**Highlights**

- Classic models of Moon formation based on the giant impact model are inconsistent with lunar geochemistry
- We provide an alternative hypothesis in which the Moon is formed directly from terrestrial silicate-rich material based in a nuclear explosion at Earth's core-mantle boundary
- Our model is consistent with the chemical composition of the Moon and satisfies the dynamic properties (energy and angular momentum) of the Earth-Moon system

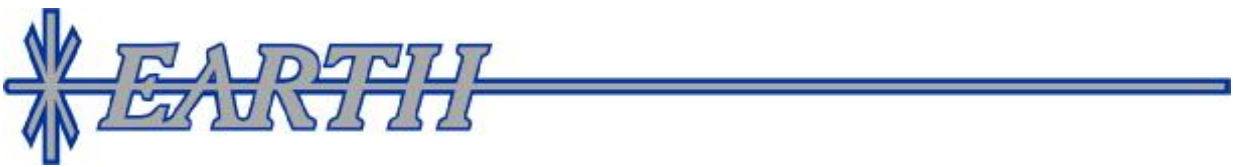

# 1. Introduction

Data from new missions by both traditional and new space faring nations, and new measurements on Apollo-era samples and lunar meteorites are revolutionizing our knowledge of the Moon. Although many of the 'classic' views of the composition and properties of the surface, interior, and atmosphere of the Moon, and their evolution through time have changed in light of these new data, the prevailing model of the formation of the Moon through a giant impact (Hartmann and Davis 1975; Cameron and Ward 1976) continues to be virtually universally adhered to.

This is perhaps surprising, because a wide range of recent studies shows that our best estimate of lunar bulk chemistry is inconsistent with dynamical models of giant impacts that reproduce the current physical properties and dynamics of the Earth-Moon system. Such models predict the chemical composition of the Moon to differ significantly from that of the Earth. From a chemical point of view, an alternative Moon formation hypothesis that is much easier to defend would result on a Moon that is simply composed of terrestrial silicate material with an isotopic composition equivalent to that of the Bulk Silicate Earth (BSE) for most elements. In this paper, we outline one such model. Our main aim is not to convince readers of the validity of our alternative hypothesis (although that would be nice); our goal is to convince readers that (a) the classic giant impact model is facing serious problems in light of a growing body of increasingly sophisticated chemical analyses and dynamical simulations and (b) alternative models should be developed and tested.

# 2. Discrepancy between dynamical models and geochemical observations

Measurements of the oxygen (Clayton and Mayeda, 1996; Wiechert et al., 2001), chromium (e.g. Shukolyukov and Lugmair, 2000; Trinquier et al., 2008), titanium (Leya et al., 2008; Zhang et al., 2012), potassium (Humayun and Clayton, 1995), and silicon (Georg et al., 2007; Savage et al. 2010; Armytage et al., 2011; Fitoussi and Bourdon, 2012) isotopic composition of lunar rocks show that the Moon and the bulk silicate Earth (BSE, i.e. mantle + crust) show a very high degree of similarity. Figure 1a shows that the oxygen isotopic compositions of Apollo-era lunar samples are indistinguishable within uncertainty from the terrestrial fractionation trend (Wiechert et al., 2001). Data in Figure 1b show that the silicon isotopic composition of the two bodies is the same within the measurement uncertainties as well (Savage et al. 2010; Armytage et al., 2011). Furthermore, the silicon isotopic composition of the bulk silicate Earth and the Moon differs from the silicon isotopic composition of chondritic meteorites, generally regarded as the building blocks for the terrestrial planets (Armytage et al., 2011; Fitoussi and Bourdon, 2012). As oxygen and silicon are the two most abundant elements on the Earth and the Moon, their identical isotopic compositions provide first-order boundary conditions for Moon-formation models.

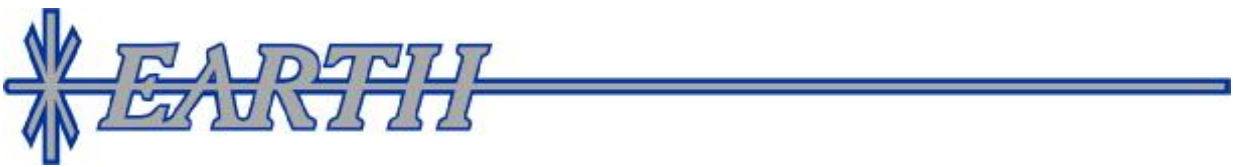

Compositional analyses of lunar meteorites (Warren, 2005) and high-precision isotopic ratio measurements on short-lived radionuclide systems such as Hf-W (Touboul et al., 2007, 2009; Münker, 2010) and Sm-Nd (Boyet and Carlson, 2007) reinforce the notion (e.g. O'Neill, 1991 and references therein) of a very high correspondence between BSE and lunar rock compositions. Figure 2a shows that the trend of $^{142}$Nd versus Sm/Nd ratio for lunar samples is consistent with their being derived from the same material as the bulk silicate Earth (Boyet and Carlson, 2007), whereas it is inconsistent with their being formed directly from undifferentiated chondritic meteoritic starting material. Similarly, figure 2b shows that the tungsten isotopic composition of lunar samples (which is set by the timing and kinetics of metallic core segregation) is identical to that of the bulk silicate Earth (Touboul et al., 2007, 2009). Finally, recent measurements of the water content of primitive lunar glasses show that even the water content of the interior of the Moon could be as high as that of the Earth's mantle (Saal et al., 2008; Hauri et al., 2011).

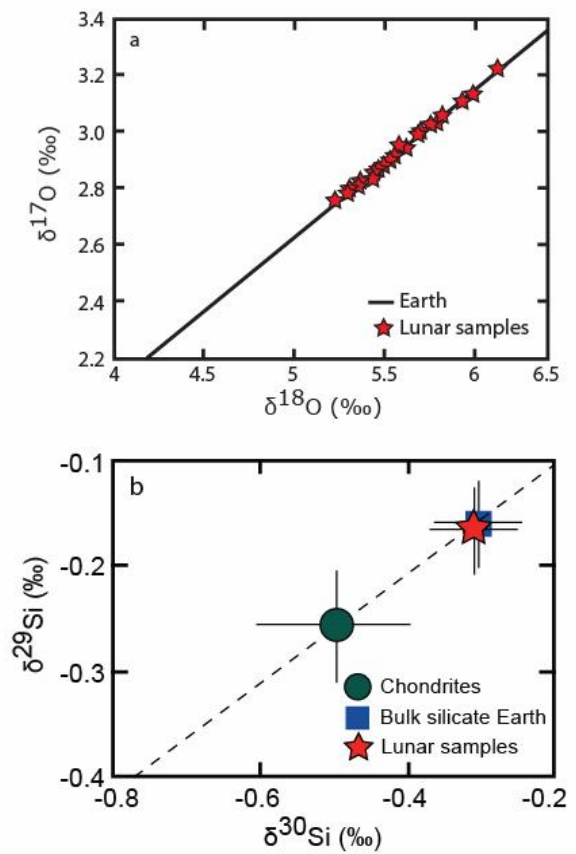

Figure 1. (a) Relation between oxygen isotopic ratios $^{17}$O/$^{16}$O and $^{18}$O/$^{16}$O (expressed as deviations in per mil from the VSMOW standard using the delta notation) for terrestrial samples (solid line) and lunar samples (symbols). Data from Wiechert et al. (2001). Error bars (2σ) are comparable to symbol size. (b) Relation between silicon isotopic ration $^{29}$Si/$^{28}$Si and $^{30}$Si/$^{28}$Si (expressed as deviations in per mil from an international rock standard using the delta notation) for chondritic meteorites, bulk silicate Earth, and lunar samples. Error bars (2σ) shown as thin lines. Data taken from Savage et al. (2010) and Armytage et al. (2011), differing slightly in absolute values from the original Georg et al. (2007) study.

These results are very hard to reconcile with the widely accepted giant-impact model for the formation of the Moon (Hartmann and Davis, 1975; Cameron and Ward, 1976). The giant impact model provides explanations for many of the first-order properties of the Earth-Moon system. The material thrown into Earth orbit after the collision consists mainly of silicate rather than metal, consistent with the silicate-rich, metal-poor chemical composition of the Moon.

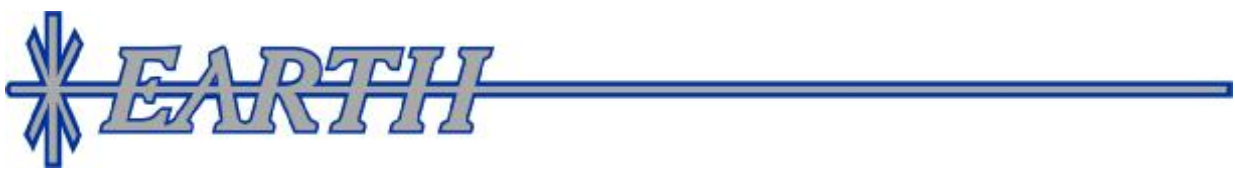

Semi-quantitative constraints on the main variables in the giant impact model (e.g. angle and relative velocity of the impact, size of the impactor) are derived from high-resolution smooth-particle hydrodynamic (SPH) simulations (e.g. Canup and Esposito, 1996; Canup, 2004, 2008). In detail the vast majority of these simulations overestimate the resulting angular momentum of the Earth-Moon system by 10-20%. At present, the 'best' angular momentum match is obtained for collisions involving a retrograde rotating Earth (Canup, 2008). Overall, a 'glancing blow' collision between the Earth and a Mars-sized impactor is consistent with the relatively large angular momentum of the Earth-Moon system.

Regardless of the collision parameters, all successful simulations indicate that by mass approximately 80% of the Moon would originate from the impactor, with only 20% originating in the Earth (e.g. Canup, 2008). Models of solar-system evolution show that it is highly unlikely for the chemical composition of the Earth and impactor to be identical (e.g. Clayton, 1993; Pahlevan and Stevenson, 2007). The giant impact model thus predicts a Moon with a chemical and isotopic composition distinctly different from that of the silicate Earth. This model prediction is inconsistent with geochemical observations (Figures 1 and 2).

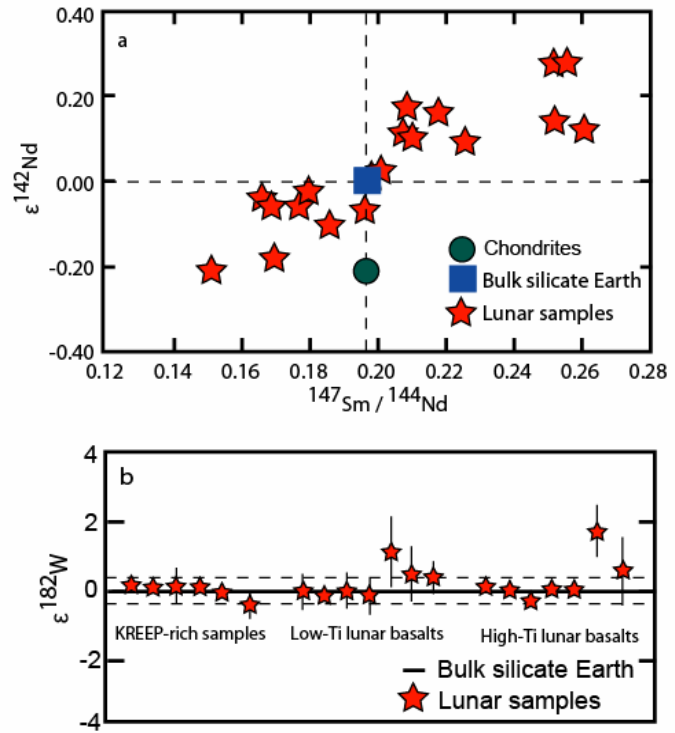

Figure 2. (a) $^{142}$Nd isotopic compositions versus Sm/Nd ratio of lunar samples compared to the values for the bulk silicate Earth and chondritic meteorites. Data compilation from Boyet and Carlson (2007), including measurements from Nyquist et al. (1995) and Rankenburg et al. (2006) . Chondrite data point taken from Boyet and Carlson (2007). Note that the lunar trend overlaps with the value for the bulk silicate Earth, not with the chondritic value (b) $^{182}$W isotopic composition (in epsilon units) of various lunar samples compared to the average of bulk silicate Earth measurements (compilation of data from Touboul et al., 2007; only data with 2σ uncertainties < 2 epsilon units are included).

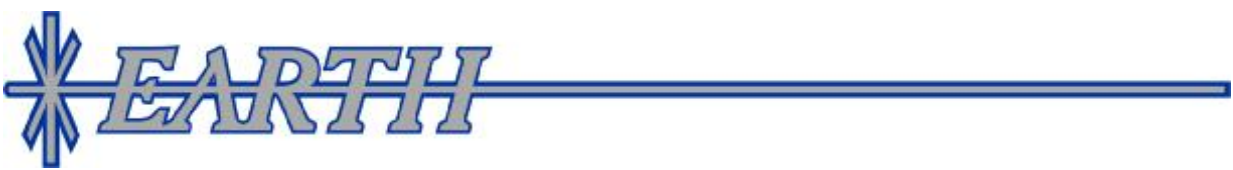

This discrepancy between dynamical models and geochemical data can not be resolved by proposing that the proto-Earth and the impactor formed at a similar distance from the Sun, for example in one of Earth's Lagrange points as proposed by Belbruno and Gott (2005). The required size difference between Earth and impactor would lead to major differences in pressure-temperature-time conditions for core formation in these bodies (which predates Moon formation in all giant impact models). The resulting differences in Si isotope and Hf-W systematics (e.g. Nimmo et al., 2010) would be detectable with current analytical techniques, even in the absence of O isotope variations, but they are simply not observed.

Resolution of this major issue without changing the main premises of the giant impact model requires total isotopic homogenisation of Earth and impactor material following the impact. Turbulent exchange between partially molten and vaporised Earth and Moon shortly after the impact has been invoked to explain the similarity in O isotopes (Pahlevan and Stevenson, 2007). The effectiveness and dynamics of this mechanism are contested (Zindler and Jacobsen, 2009; Melosh, 2009), and follow-up studies of the initial Pahlevan and Stevenson (2007) mechanism have highlighted several serious problems for the post-impact equilibration hypothesis.

First, even if this process could explain the O isotopic similarity, it is highly unlikely that such a mechanism can also fully homogenise initial differences in the isotopic compositions of much heavier, refractory elements including Si, Cr, Ti, Nd, Hf and W. Recently, Pahlevan et al. (2011) estimated the implications of a turbulent exchange equilibration scenario for silicon isotope variations. Their model predicts that equilibration in oxygen isotopes should be accompanied by a concentration of lighter silicon isotopes in lunar material compared to the Earth, something that again is not observed (Figure 1b).

Second, although full equilibration between the proto-Moon and the full silicate Earth is required, SPH simulations suggest that impactor material may form a stable layer covering several hundred kilometres of the surface of the Earth after the impact, preventing equilibration of the orbiting proto-lunar material with the Earth (Nakajima and Stevenson, 2012). Third, detailed studies of the accretion process of lunar materials from the disk surrounding the Earth after a giant impact show that only the final one-third of the mass accreting to the Moon originates from a location in the disk that is close enough to Earth to enable equilibration (Salmon and Canup, 2012). Salmon and Canup (2012) suggest that the Moon could hence be covered with a veneer of equilibrated material with terrestrial isotopic compositions, 'hiding' the impactor material in the deeper subsurface of the Moon. However, the Apollo sample collection includes volcanic lunar samples originating at least hundreds of kilometres below the lunar surface (e.g. Grove and Krawczynski,

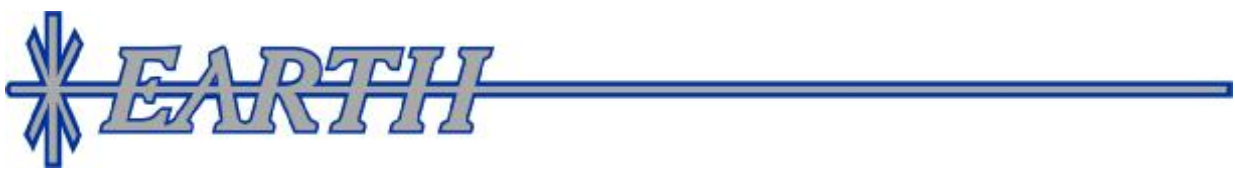

2009). These also show isotopic compositions that are identical to those of the Earth, inconsistent with expectations from the Salmon and Canup (2012) model. Given this significant and expanding number of inconsistencies produced by current versions of the giant impact model in light of geochemical data, alternative hypotheses for lunar formation should be explored (e.g. Zindler and Jacobsen, 2010; Melosh 2009).

The observed similarities in composition between Moon and BSE prompted us to revisit the hypothesis that the Moon was formed directly from terrestrial mantle material, as first proposed in the 'fission' hypothesis (Darwin, 1879). Fission is without a doubt the formation process that can most easily explain compositional similarities between Earth and Moon (e.g. Zhang et al., 2012; Cuk and Stewart, 2012). In the Darwin hypothesis, the Moon originated from a hot, deformed and fast-spinning Earth. In Darwin's model, the centrifugal forces marginally exceeded the equatorial attraction, and the Moon was formed from resonant effects of solar tides. In the beginning of the 20$^{th}$ century Moulton (1909) and Jeffreys (1930) showed that solar tidal frictions limited the height of any terrestrial tidal bulge and could not lead to the process Darwin had proposed.

Subsequently, Ringwood (1960) and Wise (1963, 1969) updated Darwin's hypothesis by including models for the thermal evolution and internal differentiation history of the Earth. They suggested that core-mantle differentiation led to a reduced moment of inertia of the Earth and hence to an increased angular velocity. The starting point for these modified models is a proto-Earth that is rotating rapidly (rotation period of ~ 2.7 h) with gravitational forces at the Earth's surface only barely exceeding centrifugal forces. In this situation, a slight increase in angular velocity would allow part of Earth's equatorial mass to be ejected into space.

The main problem with the resulting so-called Darwin-Ringwood-Wise (DRW) model was and remains the fact that the current Earth-Moon system possesses an angular momentum that is only ~ 27% of that required for a 2.7 h rotation rate of the proto-Earth. In the absence of viable models for a decrease in angular momentum by a factor of ~4 during Moon formation and subsequent Earth-Moon system evolution, the 'fission' hypothesis was abandoned before the first lunar sample return by the Apollo 11 mission.

In this paper we re-examine the dynamics of the Earth-Moon system and the energetics of initial Earth-Moon separation. In contrast to previous 'fission' models, our conservative assumption is that the angular momentum of the proto-Earth before Moon formation is close to that of the present-day Earth-Moon system. This is in full agreement with assumptions made in recent three-dimensional hydrodynamic simulations of a giant impact origin for the Moon (Canup, 2008). We estimate the amount of energy that is required to separate Earth and Moon in this case and propose that nuclear fission is the only known natural process that could supply this missing energy in a few milliseconds (Seifritz et al., 2012).

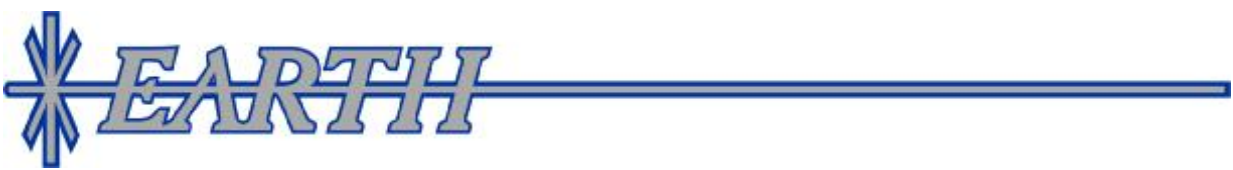

We then show that it is feasible to form the Moon through the ejection of terrestrial silicate material triggered by a nuclear explosion at Earth's core-mantle boundary (CMB), causing a shock wave propagating through the Earth. Hydrodynamic modelling of this scenario (Anisichkin at al., 1999; Voronin, 2007, 2011) shows that a shock wave created by a rapidly expanding plasma resulting from such an explosion disrupts and expels overlying mantle and crust material. This can result in the formation of a Moon-sized silicate body in Earth orbit. The energy required for this to occur, although dependent on a wide range of poorly constrained variables, is well within the range that can be produced by a nuclear explosion. Our hypothesis straightforwardly explains the chemical similarities between Earth and Moon, and connects Moon formation with processes that took place following Earth's early internal differentiation. Unlike previous 'fission' models (Ringwood, 1960; Wise, 1963, 1969) it does not rely on assuming an initial angular momentum of the Earth-Moon system that is much higher than presently observed.

**3. Earth-Moon dynamics**

Consider the Earth and Moon as a gravitationally bound two-body system, where in the ground state, just prior to separation, the Moon is a part of the Earth. In the excited state, after separation, the Moon circles the Earth at a distance, $r_{EM}$, which we consider as a variable. From laser-reflection data it has been established that $r_{EM}$ currently increases by approximately 4 cm per year (Chapront et al., 2002; Williams and Dickey, 2002). Linear extrapolation backwards in time to 4.5 Ga ago yields $r_{EM} \approx 2*10^8$ m. Although the rate of increase in $r_{EM}$ was lower in Precambrian times (e.g. Williams, 1997), the actual value of $r_{EM}$ is expected to be considerably smaller because the separation is an effect of tidal interaction, and tidal forces (proportional to the third power of the distance) were considerably stronger shortly after separation. Hence we will consider the upper value of $r_{EM}$ to be $1*10^8$ m. Contrary to the present situation, at these distances the gravitational force on the Moon exerted by the Earth dominates over the attraction exerted by the Sun by a factor of two or more.

The gravitational force, $F_g$, between Moon (mass $m_M$) and Earth ($m_E$) can be expressed by:

$$F_g = -g \frac{m_M m_E}{r_{EM}^2}, \tag{1}$$

where γ is the gravitational constant. The corresponding gravitational potential energy, $E_g$, for this two-body system becomes:

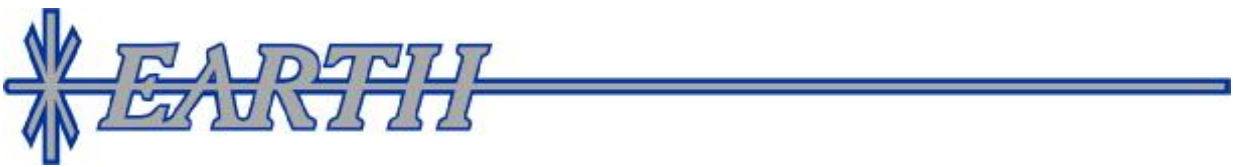

324 $$E_g = -g \frac{m_M m_E}{r_{EM}}. \tag{2}$$

325 with the assumption that $E_g = 0$ at infinite distance.

326 The first Kepler law states that the orbit of an orbiting planet/moon is an ellipse. In our first-
327 order approach we take a particular case, namely the circle, and according to the second Kepler law
328 $F$ will point to the centre of the circle. We attach the reference frame to the centre of the body with
329 $m_E$, where $m_E \gg m_M$ such that we may replace the reduced mass of the system by $m_M$. In this case
330 the centripetal force may be written as:

331 $$F_c = \frac{m_M v^2}{r_{EM}}, \tag{3}$$

332 where $v$ is the velocity of $m_M$ relative to the centre of mass.

333 Denoting the rotational motion of the two bodies by their moments of inertia, $I$, and the
334 rotational frequency, $\omega$, we may write for the sum of the rotational and potential energy, $E_{tot}$, of the
335 two body system:

336 $$E_{tot} = \frac{1}{2}(I_M w_M^2 + I_E w_E^2 + m_M v^2) - g \frac{m_M m_E}{r_{EM}}. \tag{4}$$

337 Since in a stationary orbit, either circular or elliptic, $F_g + F_c = 0$,

338 $$m_M v^2 = g \frac{m_M m_E}{r_{EM}}, \tag{5}$$

339 equation (4) reduces to:

340 $$E_{tot} = \frac{1}{2}(I_M w_M^2 + I_E w_E^2 - g \frac{m_M m_E}{r_{EM}}). \tag{6}$$

341 The total angular momentum of the system, $L$, is according to Steiner's theorem given by:

342 $$L = I_M W_M + I_E W_E + m_M r_{EM} v. \tag{7}$$

343 In our model we make a transition from a ground state in which the system is a single body:

344 $$E'_{tot} = \frac{1}{2} I'_E (w'_E)^2, \text{ and } L = I'_E W'_E, \tag{8}$$

345 to a two-body state with energy and angular momentum given by equations (6) and (7). In the
346 transition the total angular momentum is conserved. The energy difference between the one- and
347 two-body states follows from the expressions for the energy in equation (8) and equation (6),
348 respectively.

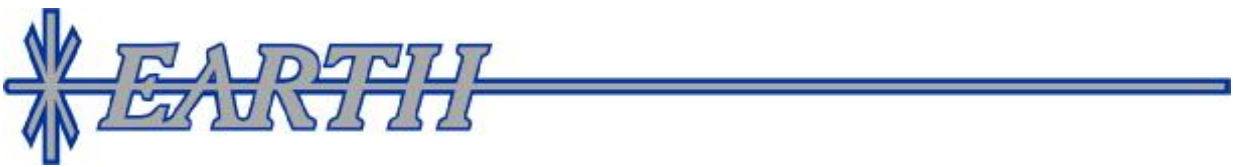

As no external torques are present in the system, the angular momentum of the total system remains unchanged during the transition. Angular momentum of the "proto-Earth" in the one-body state is redistributed between the Earth and Moon in the two-body state. The angular momentum of the present Earth-Moon system is a vectorial sum of the angular momentum due to the Moon's rotation around Earth ($2.9*10^{34}$ kg m$^2$ s$^{-1}$) in a plane with a 5.1° angle with respect to the plane of the ecliptic, and the Earth rotation around its axis ($5.9*10^{33}$ kg m$^2$ s$^{-1}$) tilted 23.5°. The resultant angular momentum of $3.5*10^{34}$ kg m$^2$ s$^{-1}$ has a tilt of 9.7°. If the proto-Earth mass equals 90 per cent of its current mass (as suggested by terrestrial accretion models, e.g. Halliday, 2004), and if the shape of the proto-Earth is assumed to be an oblate ellipsoidal with a longer axis twice as long as the shorter axes, angular momentum conservation leads to an Earth rotation period of 5.8 h ($w'_E = 3*10^{-4}$ s$^{-1}$). This rotation period is not as extreme as one may think. At present, Jupiter (which in view of its mass is not likely to have been affected by impacts from smaller bodies) has a rotation period of < 10 h, and Saturn ~12 h. Even shorter (< 5 h) proto-Earth rotation periods are currently considered to be realistic starting points for giant impact dynamical model simulations based on significant loss of angular momentum of the Earth-Moon system from the time just before Moon formation until the present day (e.g. Canup, 2012; Ćuk and Stewart, 2012).

The radius of the orbit and the energy required for the transition are coupled since angular momentum conservation is imposed. Figure 3 shows the sum of the rotational and potential energy calculated using equation (6) for a two-body system with an angular momentum equal to the present-day value for the Earth-Moon system, as a function of the initial distance between Earth and Moon. The figure shows a maximum at a distance of $1.8*10^7$ m, indicating that if lunar-orbit radius exceeds this value, the Moon is unable to return to the ground state. The energy, $E$, required to bring one Moon mass from the ground state to beyond $1.8*10^7$ m is given by

$$E = \int_{r_0}^{r_{max}} F dr \tag{9}$$

where $r_0$ and $r_{max}$ are the Moon position in proto-Earth and the distance where the maximum occurs in Figure 3. The force $F$ is the vectorial sum of the centrifugal force on the Moon in its ground state and the gravitational force. Substitution in equation 9 and subsequent integration leads to:

$$E = m_M (w'_E)^2 r_0 (r_{max} - r_0) + g m_M m_E \left(\frac{1}{r_{max}} - \frac{1}{r_0}\right) \tag{10}$$

The energy of the ground state is indicated in Figure 3 by a circle on the corresponding curve. Assuming $r_0 = 5*10^6$ m (i.e. the future lunar material positioned near the Earth's surface), the ground state energy for $L = L_p$ is $2.87*10^{30}$ J and consequently the energy required to reach a distance of $1.8*10^7$ m is $(5.37-2.87)*10^{30} = 2.5*10^{30}$ J. Figure 3 also shows the effect of changing the assumed value for the angular momentum of the proto-Earth. It is estimated that the Earth–Moon system angular momentum decreased by a few to ten per cent since the Moon was formed ($1.0\ L_p < L < 1.1\ L_p$, Canup, 2008) due to gravitational interaction with the Sun. Figure 3 shows the sum of the rotational and gravitational energy for $L$ values ranging from 0.9 to 1.1 times that of the present. The critical Earth-Moon distance at which the total energy reaches a maximum decreases from $2.0*10^7$ m to $1.7*10^7$ m as $L$ is increased from 0.9 to 1.1 $L_p$. The corresponding energy that needs to be added to reach these distances decreases slightly from $2.6*10^{30}$ J to $2.4*10^{30}$ J. Overall, the energy required is therefore not very sensitive to the assumption on the extent of the conservation of angular momentum within these tight bounds. We note that if the angular momentum of the Earth prior to Moon formation was much higher than the angular momentum of the Earth-Moon system today (e.g. Canup, 2012, Cuk and Stewart, 2012), the required energy would decrease significantly, and would approach zero as the rotation period of the Earth approaches ~2.3 h (equal to its stability limit). For the remainder of this study we assume $2.5*10^{30}$ J as the maximum additional energy required to bring one lunar mass into Earth orbit.

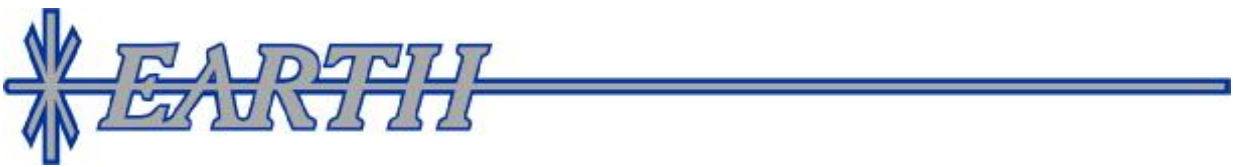

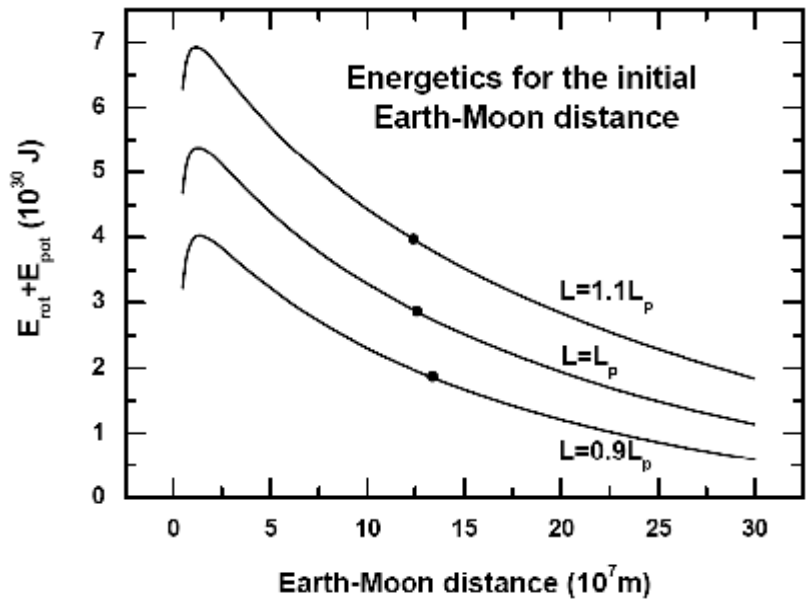

Figure 3. Energetics of the initial Moon orbit in a Earth-Moon system with conserved angular momentum assuming angular momentum $L$ same as today ($L_p$), 0.9 x today, and 1.1 x today. Circles on curves mark positions at which the energy of the excited Earth-Moon state exceeds the energy of the ground state in which the Moon is a part of the proto-Earth. Only the sections of the curves to the left of the circles mark the physically relevant region.

To illustrate the implications of this model, if we assume the maximum $r_{EM}$ value of $1*10^8$ m and $L = L_p$, the gravitational potential energy $E_g$ has a value of $-1.2*10^{29}$ J and the corresponding velocity of the proto-Moon becomes, according to equation (5), $1.9*10^3$ m s$^{-1}$. This corresponds to a rotation period of 92 h or ~3.8 d around the proto-Earth and an angular momentum carried by the proto-Moon of $1.3*10^{34}$ kg m$^2$ s$^{-1}$, which would correspond to approximately 40% of the total angular momentum $L_p$ of the Earth-Moon system. Immediately after the separation of the proto-Earth and proto-Moon, the rotation period of the proto-Earth would become 9.0 h. Due to tidal forces, energy and angular momentum are transferred from the Earth to the Moon until they have, with their increased mass, the present properties.

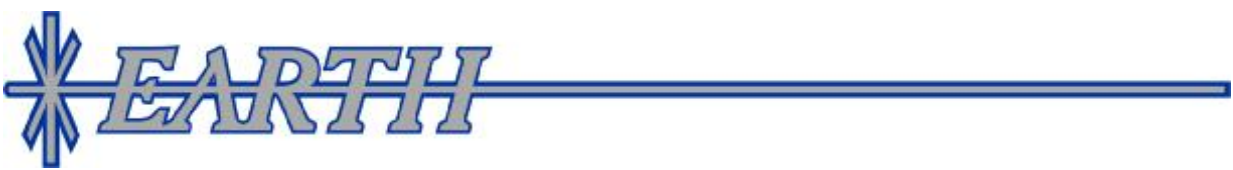

**4. Georeactors at the core-mantle boundary**

In the previous section we showed that in the conservative case of the angular momentum of the proto-Earth being close to that of the Earth-Moon system today, the dynamics of the formation of the Moon from terrestrial materials requires ~ $2.5*10^{30}$ J rapidly generated additional energy. To our knowledge, the only realistic known process that can generate this amount of energy in a very short time window, in the aftermath of large-scale differentiation processes, is nuclear fission. Here, we consider the possibility that the Moon was formed from the ejection of terrestrial mantle material triggered by a shock wave generated by a nuclear explosion of a natural nuclear reactor (a georeactor) at Earth's core-mantle boundary (CMB).

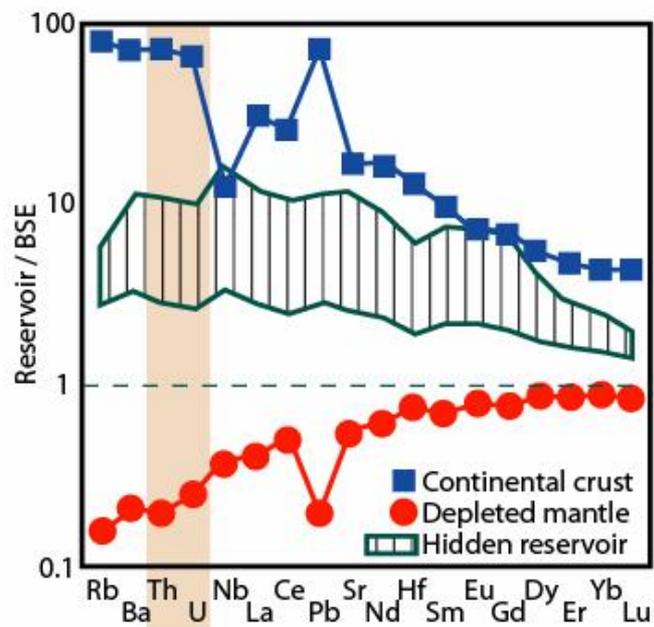

We (de Meijer and van Westrenen, 2008) recently assessed in detail the feasibility of georeactors (Herndon, 1992; Hollenbach and Herdon, 2004) in Earth's CMB. Only the main points of that paper are summarised here. For details see de Meijer and van Westrenen (2008). In our model, georeactors are a natural consequence of concentrating significant proportions of our planet's U and Th budgets in the CMB region, as suggested by several current geochemical models of Early Earth evolution (see Boyet and Carlson, 2005; Tolstikhin and

Figure 4. . Recent estimates for the lithophile element concentrations in a solid silicate 'hidden reservoir' in the core-mantle boundary region (hatched field), compared to concentrations in the continental crust and the depleted mantle (simplified after Carlson and Boyet, 2009). Upper and lower limits of the hatched field assume the volume of the hidden reservoir to be equivalent to 4 and 26% of the Bulk Silicate Earth (BSE), respectively. Continental crust data from Rudnick and Gao (2003), depleted mantle values from Boyet and Carlson (2006). All concentrations are relative to abundances in the Bulk Silicate Earth (BSE).

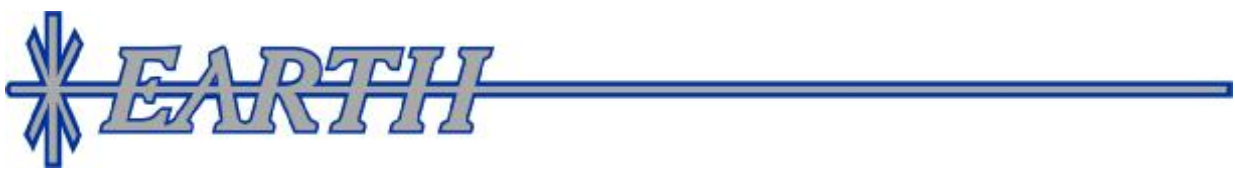

Hofmann, 2005; Tolstikhin et al., 2006). Boyet and Carlson (2005) showed that terrestrial mantle samples have an enhanced Sm/Nd ratio compared to the undifferentiated meteorites that are assumed to be the Earth's building blocks. As both Sm and Nd are lithophile elements one expects no difference between the Sm/Nd ratio of mantle samples and undifferentiated meteorites. The mantle should therefore contain a complementary reservoir, with a low Sm/Nd ratio, which has remained isolated from the rest of the mantle since its formation: a so-called 'hidden reservoir'.

Because of efficient mantle mixing through convection processes, the only viable option for the location of this reservoir is the core-mantle boundary region. In addition to being enriched in Nd with respect to Sm, the CMB region must also be enriched in other lithophile elements including uranium and thorium. Recent estimates from Carlson and Boyet (2009), reproduced in Figure 4, show that CMB region concentrations of uranium and thorium are approximately two orders of magnitude higher than values for the 'regular' depleted mantle. The same likely held for plutonium at the time of Moon formation.

In addition to this general enrichment, the mineralogy of the lowermost mantle provides an extra concentration step for uranium, thorium and plutonium (de Meijer and van Westrenen, 2008). High-pressure experiments on putative mantle rocks, in combination with geophysical observations of the density structure of Earth's interior, suggest that the mineralogy of the CMB region is relatively simple, consisting of magnesium-iron silicate perovskite (MgPv), ferropericlase, and calcium silicate perovskite. Although it is currently impossible to obtain U and Th distribution data at CMB conditions (i.e. pressures of approximately 125 GPa and temperatures of 2500–4000 K), experiments at 25 GPa and 2600 K indicate that U and Th concentrations in calcium silicate perovskite are 3–4 orders of magnitude greater than concentrations in co-existing MgPv (e.g. Hirose et al., 2004; Corgne et al., 2005). The recently discovered new high-pressure form of MgPv, named postperovskite (Murakami et al., 2004) which may be stable in the CMB, is unlikely to influence this distribution. Ferropericlase generally incorporates even lower concentrations of trace elements than MgPv (e.g. Walter et al., 2004).

De Meijer and Van Westrenen (2008) calculate that selective incorporation of fissionable material by calcium silicate perovskite (CaPv) leads to concentrations of > 4 ppm U, ~8 ppm Th, and 19 ppb of $^{244}$Pu in CaPv in the CMB region at the time of Moon formation (50-150 Ma after solar system formation, e.g. Touboul et al., 2007). In the absence of water, this is a factor of fifteen to twenty lower than required for igniting and maintaining a nuclear breeder reactor if U, Th and CaPv are assumed to be distributed homogeneously throughout the CMB (de Meijer and van Westrenen, 2008).

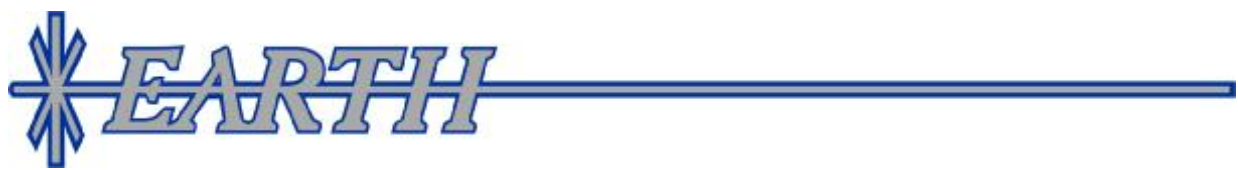

**5. Nuclear excitation**

The calculations above show that without additional concentration factors, U, Th and Pu concentrations in the CMB are insufficient to reach criticality. Additional concentrations can be achieved by a combination of two processes: growth of the relative concentration of the fissile materials by a transient pressure wave, induced by an impact at the Earth's surface (Anisichkin, 1997; Voronin and Anisichkin, 2001), and/or the development of compositional heterogeneities (de Meijer and Van Westrenen, 2008).

As shown by Voronin (2011) for fissile material at the CMB, an impact of a 100km-diameter asteroid can create a transient pressure increase of several TPa at the CMB, sufficient to concentrate fissile material from a subcritical to a supercritical condition followed by a nuclear explosion. Regarding the development of compositional heterogeneities, it should be noted that small-scale heterogeneities exist in the core-mantle boundary region even today (e.g., van der Hilst et al., 2007): volumes exhibiting both higher-than-average and lower-than-average wave propagation speeds, with diameters as small as 30 km, are now resolvable. Some studies suggest that the bottom of the mantle is partially molten today, forming a so-called 'basal magma ocean' (e.g. Williams and Garnero, 1996; Labrosse et al., 2007; Lee et al., 2010).

Although the precise nature and composition of these heterogeneities remains unresolved, this suggests that significant local concentration factors, in addition to the general CMB and CaPv enrichments described above, are entirely plausible even today. The dynamics of the CMB 4.5 Ga ago are poorly explored. The higher rotation rate of Earth at that time, and higher interior temperatures, are likely to have facilitated local concentration of density heterogeneities to levels that exceed those currently observed, due to centrifugal forces and buoyancy effects associated with local heating.

A combination of impact-induced densification and compositional heterogeneity make a concentration factor of fifteen to twenty compared to the fully homogeneous scenario not unreasonable (de Meijer and van Westrenen, 2008). In the next section we illustrate how the Moon can be formed in the aftermath of a nuclear explosion.

**6. Moon formation**

Reactor physics calculations on the excursion of a georeactor (Seifritz et al., 2012) indicate that the nuclear energy is released in a few milliseconds, creating a plasma with temperatures on the order of $10^{10}$ K and resulting in a shock wave. Anisichkin at al. (1999) and Voronin (2011) simulated the effects of their propagation through the silicate Earth. In these simulations energy and angular momentum are conserved. Figure 5 shows the time evolution of one of their hydrodynamic simulations. In this particular case, supercriticality of a CMB reactor is achieved by an impact of a 100km-diameter asteroid (body **1** in Figure 5) hitting a rapidly rotating differentiated Earth (with an equatorial radius of 7000 km) at the equator with a velocity of 30 km s$^{-1}$ (Figure 5a).

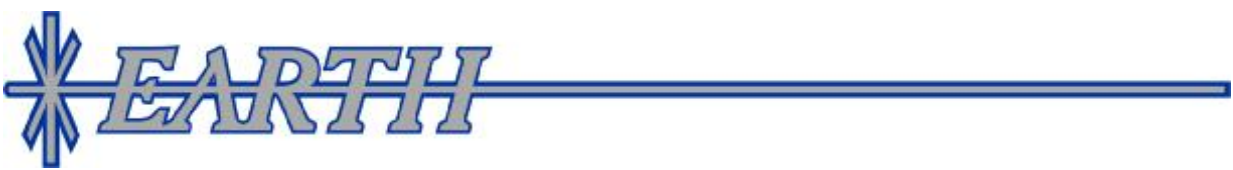

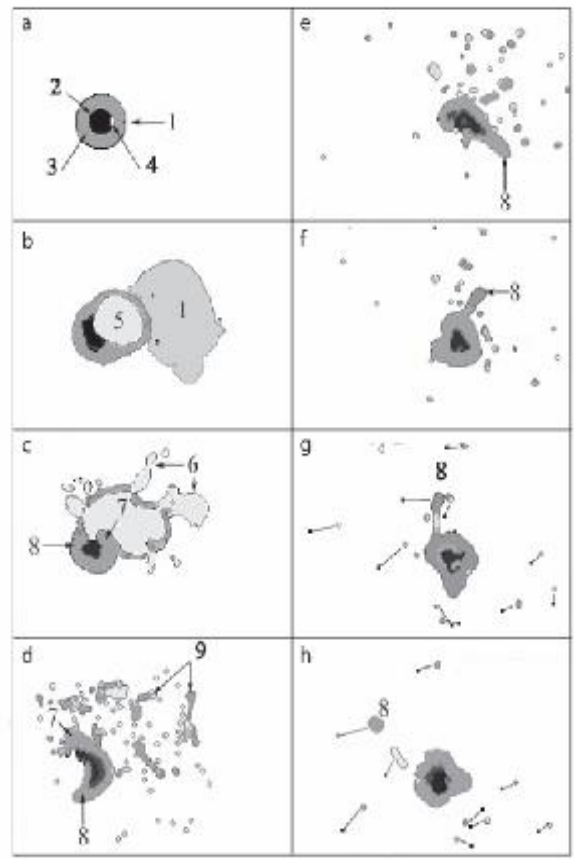

In Figure 5c (~40 minutes after impact) the plasma and shock wave are shown to fragment the Earth's mantle and crust, with jets of plasma escaping to space. Approximately 1 hour after impact, fragments of crust and mantle are ejected into orbit (Figures 5def). In this particular simulation, the Moon (fragment **8** in Figures 5cd) is still part of the remaining Earth at this stage. The Earth returns to a more spherical shape with the Moon attached by a thin 'neck' (Figure 5g), which detaches from the Earth approximately 3 hours after the impact-triggered excursion (Figure 5h). Other fragments return to Earth or are lost to space depending on their energy and angular momentum. The final Earth:Moon mass ratio in this particular simulation agrees with observation, and the Moon is essentially fully comprised of terrestrial silicate material.

Figure 5. Snapshots of hydrodynamic simulations of Moon formation (Voronin 2011). **1** – asteroid impacting on Earth's surface; **2** – Earth's core; **3, 7, 9** – mantle; **4** – location of nuclear explosion; **5, 6** – explosion products / plasma; **8** - separating Moon-like silicate-rich fragment**.**

The Anisichkin (1999) and Voronin (2007, 2011) models provide a proof-of-concept of the Moon formation scenario we propose. The required fission energy depends on the assumptions made in the simulation. For example, the required energy increases with decreasing rotation speed of the Earth, decreasing equatorial radius, and increasing mass of the proto-Earth. Clearly, at this stage, only a limited number of hydrodynamic simulations of this scenario have been conducted, and large parts of parameter space remain to be explored. For example, the recent study by Ćuk and Stewart (2012) mentioned above suggests that the Earth-Moon system loses a significant portion of its angular momentum shortly after its formation, due to resonances between the Moon, the Earth's core, and the Sun. If this is correct, the angular momentum constraint on lunar formation models is too conservative and may be significantly relaxed.

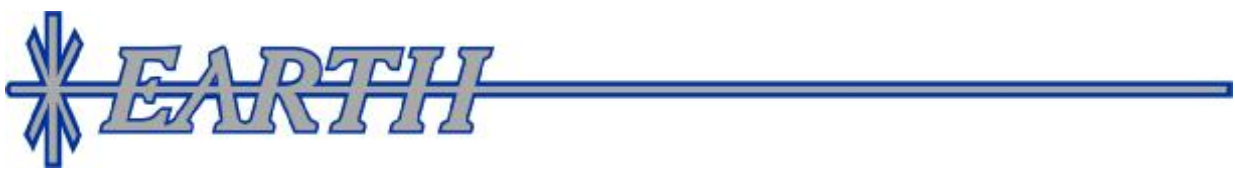

The end member models we have discussed above provide estimates for the range of fission energy required to form the Moon. Hydrodynamic models leading to the formation of the Moon as shown in Figure 5, starting with a 7000 km radius Earth with a fast 3 h rotation period, require a minimum fission energy of $0.6*10^{29}$ J (Voronin, 2007, 2011). In our discussion of Earth-Moon dynamics, we assumed an initial ~6000 km radius Earth with a conservative rotation period of 5.8 h and derived a required fission energy of $2.5*10^{30}$ J. The next question to be addressed is whether the U-Th inventory of the CMB is sufficient to provide between $0.6*10^{29}$ J and $2.5*10^{30}$ J of fission energy.

**Table 1.** *Masses and isotopic abundances of Th and U isotopes in the Bulk Silicate Earth (BSE) (McDonough, 2003).*

|  | $^{232}$Th | $^{235}$U | $^{238}$U | Total mass |
|---|---|---|---|---|
| $t_{1/2}$ (Ga) | 14.05 | 0.70 | 4.47 |  |
| $m$ ($10^{17}$ kg) ($t = 0$) | 3.15 | $5.87*10^{-3}$ | 0.80 | 3.95 |
| Isotopic abundance ($t = 0$) | 100% | 0.73% | 99.27% |  |
| $m$ ($10^{17}$ kg).($t = -4.5$ Ga) | 3.94 | 0.52 | 1.62 | 6.06 |
| Isotopic abundance ($t = -4.5$ Ga) | 100% | 24.3% | 75.7% |  |

**7. Fission energy production in the CMB**

Table 1 presents the amounts of $^{232}$Th, $^{235}$U and $^{238}$U according to a commonly used Bulk Silicate Earth (BSE) compositional model (McDonough, 2003) for both the present and 4.5 Ga ago. From Table 1 one may calculate that fission of 1 kg of a natural mixture at $t = -4.5$ Ga of $^{232}$Th, $^{235}$U and $^{238}$U yielded $8.21*10^{13}$ J. Consequently, it requires fission of $7.3-320*10^{14}$ kg of the natural (U+Th) mixture to separate the Moon from the Earth at $t = -4.5$ Ga for the two models discussed in this paper. The concentration of (U+Th) to allow a georeactor to become critical is estimated to be of the order of 150 ppm (U+Th) (de Meijer and van Westrenen, 2008). Hence the corresponding mass of CMB material involved is of the order of $4.9-210*10^{18}$ kg. At a silicate rock density at CMB conditions close to $5.5*10^{3}$ kg m$^{-3}$, as derived from seismic observations (e.g. Dziewonski and Anderson, 1981), this mass corresponds to a sphere with a radius of approximately 60-210 km. Of course the shape of the reactor would not necessarily be spherical, but this calculation demonstrates that such a volume is fully compatible with our present understanding of the dimensions of a 'hidden reservoir' near the CMB.

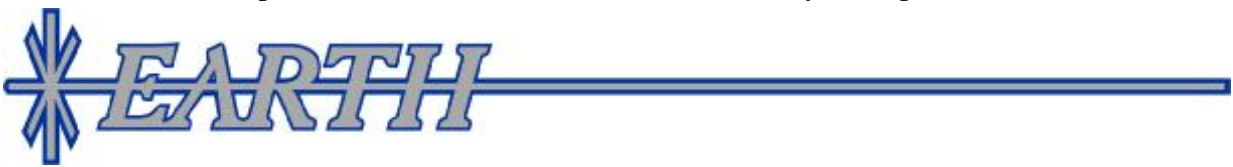

The maximum required mass of $320*10^{14}$ kg of the natural (U+Th) mixture corresponds to 5% of the fissionable (U+Th) in the BSE at $t = -4.5$ Ga. If approximately half of the BSE (U+Th) budget was concentrated in the CMB, as proposed by Tolstikhin and Hofmann (2005) and Tolstikhin et al. (2006), this corresponds to a maximum of ~ 25% of the CMB (U+Th) content. The minimum required values are ~60 times smaller than these maximum values. This range of percentages does not seem unrealistic.

**8. Timing of Moon formation**

If the Moon originates directly from the Earth's mantle, the temporal evolution of the two bodies must be intimately connected. Our hypothesis is consistent with the timescales of Earth differentiation and lunar formation. Core-mantle differentiation must have preceded Moon formation because of the relatively low iron content of the Moon. The timing of terrestrial core formation is derived from the interpretation of tungsten isotopic data for terrestrial and meteorite samples. Core segregation in the Earth is estimated to occur at $t = 30$-$50$ Ma after the start of the solar system (e.g. Kleine et al., 2002; Yin et al., 2002; Kleine et al., 2004). Based on Hf-W analyses of lunar rocks Touboul et al. (2007, 2009) conclude that the Moon was formed at $t = 50$-$150$ Ma, after completion of most of the core-mantle differentiation of Earth.

A second prerequisite for the presence of georeactors at the CMB is the availability of sufficient concentrations of uranium and thorium. As outlined above and in our previous paper (de Meijer and van Westrenen, 2008), such elevated concentrations of U and Th accompany the formation of a 'hidden reservoir' in the CMB (Figure 4). Evidence for the timing of the formation of this reservoir is provided by the $^{146}Sm/^{142}Nd$ chronometer, which points to a date for the formation of the hidden reservoir of around $t = 30$ Ma (Boyet and Carlson, 2005). Again, this is consistent with the timing of Moon formation.

The completion of core-mantle differentiation, the formation of a hidden reservoir, and the formation of the Moon all took place in a relatively narrow time-interval. In our hypothesis this sequence of events is necessarily correlated. Although giant impacts are expected towards the end of accretion of the planets in the solar system, due to the presence of many Mars-sized planetesimals on eccentric orbits at this time, the narrow time interval for these processes observed for the Earth-Moon system is more of a coincidence in that case.

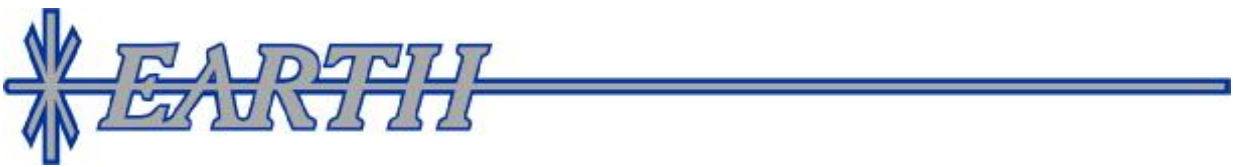

## 9. Supporting evidence

The main supporting evidence for our hypothesis is the correspondence in isotopic and elemental composition between the BSE and lunar rocks. Positive identification of nuclear fission products from the georeactor in lunar material would strongly support our hypothesis. In our preceding paper (de Meijer and van Westrenen, 2008) we quantitatively assessed a wide range of potential changes in isotopic abundances in the Earth due to georeactor activity. We showed that helium and xenon isotope ratios are the primary messengers of georeactor activity. The magnitude of changes in these ratios depends critically on the proportion of supercritical georeactor material that is eventually ejected (e.g. Figure 5), a parameter that is poorly constrained at present.

In principle, the strongest indicator would be the presence of high concentrations of $^3$He in the Moon. Unfortunately, the continuous influx of $^3$He from the solar wind means surface lunar rocks cannot be used to measure the 'indigenous' lunar $^3$He levels. Any $^3$He found at greater depths in the Moon will be a strong indication of the involvement of a georeactor in the origin of the Moon. Elevated levels of $^{136}$Xe are also expected if georeactor fission products were incorporated in the ejected material that formed the Moon. Marti et al. (1970) do report a small excess of $^{136}$Xe, which Boulos and Manuel (1971) subsequently coupled to the activity of extinct $^{244}$Pu. We have previously shown (de Meijer and van Westrenen, 2008) that the amount of $^{136}$Xe is too large for $^{244}$Pu to be the source of the surplus $^{136}$Xe in terrestrial gas wells, and that instead $^{136}$Xe is an indicator of georeactor activity. Due to the similarity in composition between Earth and Moon this argument holds even more strongly for the Moon, as, with a half-life time of 80 Ma, the $^{244}$Pu content at the time of Moon formation would already have been reduced by at least a factor of two compared to the initial terrestrial concentration.

We interpret the measured xenon isotopic composition in the lunar surface sample analysed by Marti et al. (1970) in terms of a mixture between solar wind xenon and "internal" xenon. Our estimate is based on a very crude model in which the "internal" Xe abundance and isotopic composition is the sum of xenon produced in the run-away georeactor and the xenon present in terrestrial mantle material (Lodders, 2003). The isotopic composition in the mantle material is taken from Busemann et al. (2000). Our estimate indicates that about 70% of the xenon in this lunar soil sample originates from solar wind. Analysis of soil and rock samples from greater depths will be more conclusive.

As up to 25% of the U and Th in the CMB is assumed to be involved in the run-away georeactor, one would at first glance expect a difference in the $^{235}$U/$^{238}$U and Th/U ratios between Moon and Earth, which is not consistent with observations (e.g. Tatsumoto and Rosholt 1970).

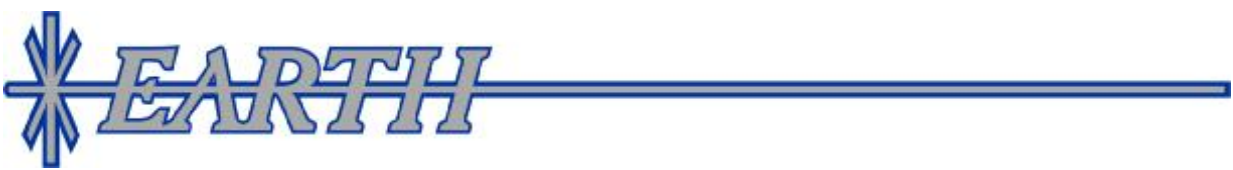

There are however a number of reasons why such differences are not likely. The main reason is that in a breeder type georeactor, both $^{235}$U and $^{238}$U (as well as $^{232}$Th) disappear by conversion to fissile materials (Anisichkin at al. 2008). The probability for interaction (mainly fission) by fast neutrons for these three isotopes is virtually identical (6.84, 7.10 and 7.00 barn at $E_n = 1$ MeV for $^{235}$U, $^{238}$U and $^{232}$Th, respectively, e.g. Rinard (1991). The actual change in the $^{235}$U/$^{238}$U or Th/U ratio will depend on the precise shape of the energy distribution of the neutrons, on the temperature and pressure at the georeactor site, and on the presence of $^{244}$Pu. To first order there will hardly be any difference of the ratios between lunar and terrestrial samples. A second factor that will reduce differences in isotopic ratios is the dilution of the georeactor material by mixing with terrestrial mantle material. As pointed out above the mass of the initial georeactor is three orders of magnitude smaller than the final mass of Moon. In summary, no significant differences in $^{235}$U/$^{238}$U or Th/U ratio are expected between lunar and Earth materials if the Moon was formed as we propose here.

**10. Conclusions**

Moon formation models have to be consistent with lunar chemistry. Current versions of the giant impact model are not. Alternative models in which the Moon is formed from terrestrial material deserve more detailed study. Here, we provide such an alternative model. We show that a nuclear explosion in the CMB can provide the missing energy source for the Darwin-Ringwood-Wise fission model for Moon formation. Our hypothesis provides a straightforward explanation for the striking similarity in elemental and isotopic composition of the Earth's mantle and lunar rocks, and is consistent with the sequence of differentiation events during our planet's earliest history. Future Moon missions returning lunar samples from greater depths may contain supportive evidence for the validity of our hypothesis. The $^3$He contents and xenon isotopic compositions in particular, would be a crucial test of this hypothesis.

**Acknowledgements:** Critical comments by Don Wise and anonymous reviewers helped clarify key aspects of this study. We are indebted to Prof. Walter Seifritz for his critical evaluation of georeactor criticality. WvW acknowledges financial support from a European Science Foundation EURYI award and the Netherlands Space Office User Support Programme Space Research.

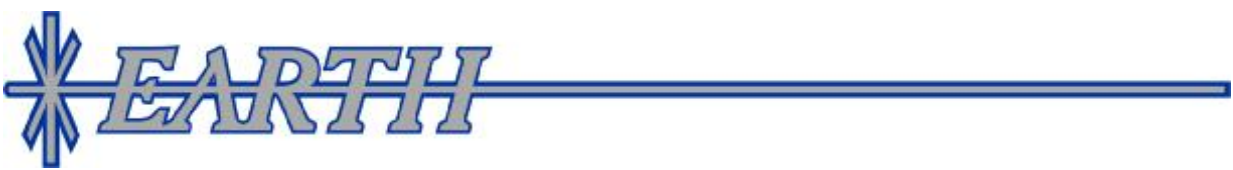

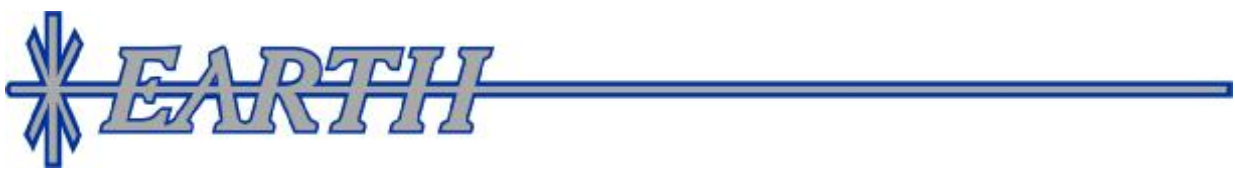

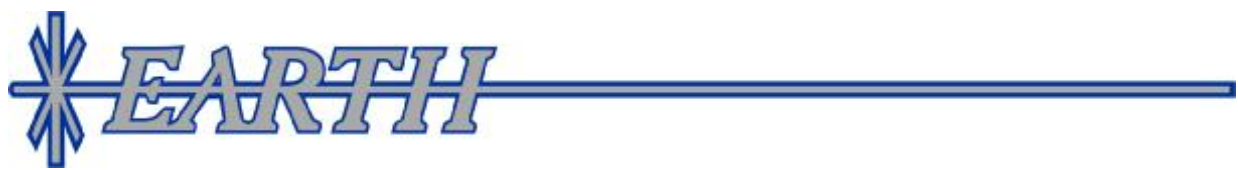

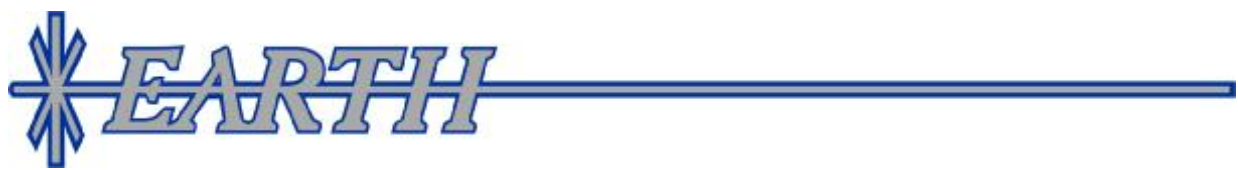

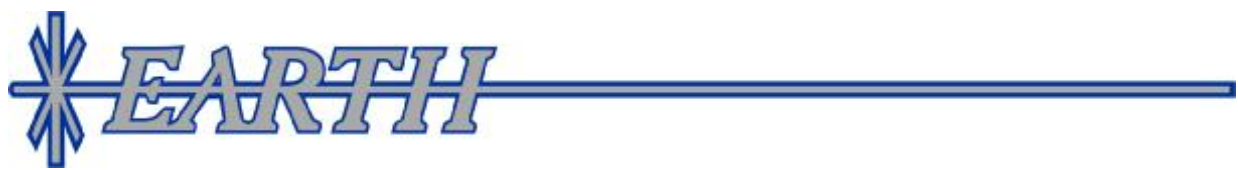

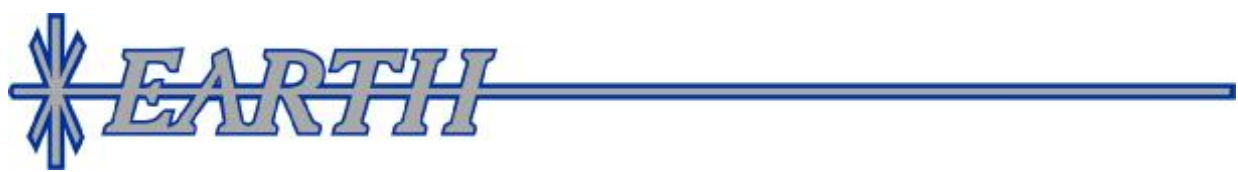

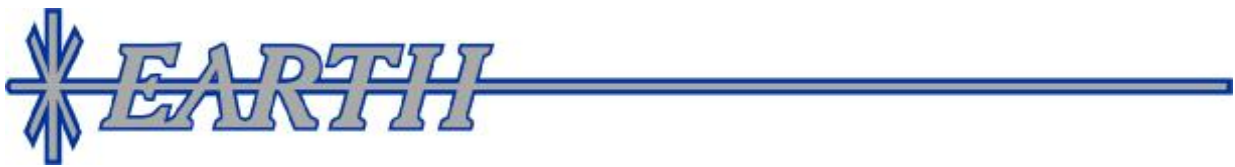